\input harvmac
\noblackbox
 
\font\ticp=cmcsc10
 
\def\Title#1#2{\rightline{#1}\ifx\answ\bigans\nopagenumbers\pageno0\vskip1in
\else\pageno1\vskip.8in\fi \centerline{\titlefont #2}\vskip .5in}

\font\ticp=cmcsc10
\font\ttsmall=cmtt10 at 8pt

\input epsf
\ifx\epsfbox\UnDeFiNeD\message{(NO epsf.tex, FIGURES WILL BE
IGNORED)}
\def\figin#1{\vskip2in}
\else\message{(FIGURES WILL BE INCLUDED)}\def\figin#1{#1}\fi
\def\ifig#1#2#3{\xdef#1{fig.~\the\figno}
\goodbreak\topinsert\figin{\centerline{#3}}%
\smallskip\centerline{\vbox{\baselineskip12pt
\advance\hsize by -1truein\noindent{\bf Fig.~\the\figno:} #2}}
\bigskip\endinsert\global\advance\figno by1}

%
%
\def\qn{quasinormal}
\def\SAdS{Schwarzschild-AdS }
\def\om{\omega}
\def\p{\partial}

\def\[{\left [}
\def\]{\right ]}
\def\({\left (}
\def\){\right )}


\lref\rasa{S. Kalyana Rama and B. Sathiapalan, ``On the role of chaos in the
AdS/CFT connection", hep-th/9905219.}
\lref\juan{J. Maldacena, Adv. Theor. Math. Phys. 2 (1998) 231, hep-th/9711200.}
\lref\witten{ E. Witten, Adv. Theor. Math. Phys. 2 (1998) 253, hep-th/9802150.}
\lref\witt{ E. Witten, Adv. Theor. Math. Phys. 2 (1998) 505, hep-th/9803131.}
\lref\gkp{S. Gubser, I. Klebanov, and A. Polyakov, Phys. Lett. B428 (1998) 105,
hep-th/9802109.}
\lref\chma{J. Chan and R. Mann, Phys. Rev. D55 (1997) 7546, gr-qc/9612026.}
\lref\price{R. Price, Phys. Rev. D5 (1972) 2419, 2439.}
\lref\mann{J. Chan and R. Mann, Phys. Rev. D55 (1997) 7546, gr-qc/9612026; 
Phys. Rev. D59 (1999) 064025.}
\lref\qnmode{The earliest papers include C. Vishveshwara, Phys. Rev. D1
(1970) 2870; W. Press, Ap. J. Lett. 170 (1971) L105. For an early review,
see S. Detweiler, in {\sl Sources of Gravitational Radiation}, L. Smarr, ed.,
Cambridge U. Press (1979) p. 221.}
\lref\schmidt{For a recent review, see 
K. Kokkotas and B. Schmidt, ``Quasi-normal modes of stars and
black holes", to appear in Living Reviews in Relativity: www.livingreviews.org
(1999).}
\lref\brady{P. Brady, C. Chambers, W. Krivan and P. Laguna,
Phys. Rev. D55 (1997) 7538, gr-qc/9611056;
P. Brady, C. Chambers, W. Laarakkers, and
E. Poison, Phys. Rev. D60 (1999) 064003, gr-qc/9902010.}
\lref\glue{C. Csaki, H. Ooguri, Y. Oz, and J. Terning, JHEP 01 (1999)
017, hep-th/9806021; R. de Mello Koch, A.Jevicki, M.Mihailescu, and
J.P.Nunes, Phys. Rev. D58 (1998) 105009, hep-th/9806125;
M. Zyskin, Phys. Lett. B439 (1998)
373, hep-th/9806128.}
\lref\gary{G. Horowitz, talk at Strings '99, Potsdam, Germany, to appear
in the proceedings.}
\lref\grla{R. Gregory and R. Laflamme, Phys. Rev. Lett. 70 (1993) 2837, 
hep-th/9301052; Nucl. Phys. B428 (1994) 399, 
hep-th/9404071.} 
\lref\hapa{S. Hawking and D. Page, Commun. Math. Phys. 87 (1983) 577.}
\lref\chop{M. Choptuik, Phys. Rev. Lett. 70 (1993) 9.}
\lref\gcd{D. Garfinkle, C. Cutler, and G. C. Duncan, gr-qc/9908044.} 
\lref\gund{ C. Gundlach, Adv. Theor. Math. Phys. 2 (1998) 1, gr-qc/9712084.}
\lref\dgm{S. R. Das, G. Gibbons, and 
S. D. Mathur, Phys. Rev. Lett. 78 (1997) 417, 
hep-th/9609052.}
\lref\clsy{E.S.C. Ching, P.T. Leung, W.M. Suen, and
K. Young, Phys. Rev. D52 (1995)
2118, gr-qc/9507035.}
\lref\bazw{A. Barreto and M. Zworski, Math. Research Lett. 4 (1997) 103.}
\lref\bulu{C. Burgess and C. Lutken, Phys. Lett. 153B (1985) 137.}
\lref\magoo{For a comprehensive review, see O. Aharony, S.S. Gubser,
J. Maldacena, H. Ooguri, and Y. Oz,
``Large N Field Theories, String Theory and Gravity",
hep-th/9905111.}
\lref\gpp{C. Gundlach, R. Price, and J. Pullin, Phys. Rev. D49 (1994) 883.}
\lref\bdhm{T. Banks, M. Douglas, G. Horowitz, and E. Martinec,
``AdS Dynamics from Conformal Field Theory", hep-th/9808016.}

%
%
\baselineskip 16pt
\Title{\vbox{\baselineskip12pt
\line{\hfil   NSF-ITP-99-70}
\line{\hfil \tt hep-th/9909056} }}
{\vbox{
{\centerline{Quasinormal Modes of AdS Black Holes}
\vskip .5cm
\centerline{and the Approach to Thermal Equilibrium}}
}}
\centerline{\ticp Gary T. Horowitz and Veronika E. Hubeny\footnote{}{\ttsmall
 gary@cosmic.physics.ucsb.edu, veronika@cosmic.physics.ucsb.edu}}
\bigskip
\centerline{\it Physics Department, University of California,
Santa Barbara, CA 93106, USA}
\bigskip
\centerline{\bf Abstract}
\bigskip
We investigate the decay of a scalar field outside a Schwarzschild anti de
Sitter black hole. This is determined by computing the complex
frequencies associated with quasinormal modes. 
There are qualitative differences from the
asymptotically flat case, even in the limit of small black holes. 
In particular, for a given angular dependence,  the decay is always
exponential - there are no power law tails at late times. In terms of the
AdS/CFT correspondence, a large black hole corresponds to an approximately
thermal state in the field theory, and the decay of the scalar field
corresponds to the decay of a perturbation of this state. Thus one obtains
the timescale for the approach to thermal equilibrium. We compute these
timescales for the strongly coupled 
field theories in three, four, and six dimensions which are dual to string
theory in asymptotically AdS spacetimes.
\Date{September, 1999}
\newsec{Introduction}

It is well known that if you perturb a black hole, the surrounding geometry
 will ``ring",
i.e., undergo damped
oscillations. The frequencies and damping times of these oscillations
are entirely fixed by the black hole, and are independent of the initial
perturbation.  These oscillations are similar to normal modes of a closed
system. However,
 since the field can fall into the black hole or radiate to infinity, 
the modes decay and the corresponding frequencies are complex. These
oscillations are 
known as ``quasinormal modes".
 For black holes in asymptotically flat spacetimes,
they have been studied for almost thirty years \refs{\qnmode,\schmidt}.
The radiation associated with these modes is expected to be seen
with gravitational wave detectors in the coming decade.
Motivated by inflation, the quasinormal modes of black holes in de Sitter space
have recently been studied \refs{\brady,\bazw}.

For spacetimes which asymptotically approach anti de Sitter (AdS) spacetime
the situation is slightly different. In the absence of a black hole,
most fields propagating in AdS
can be expanded in ordinary normal modes. The cosmological constant
provides an effective confining box, and solutions only exist with discrete
(real) frequencies. However once a black hole is present, this is no longer
the case. The fields can now fall into the black hole and decay. There 
should exist complex frequencies, characteristic of the black hole, which
describe the decay of perturbations outside the horizon. We will compute
these quasinormal
frequencies below, for spacetimes of various dimensions.

The quasinormal frequencies of AdS black holes have a direct interpretation
in terms of the dual conformal field theory (CFT) 
\refs{\juan,\witten,\gkp,\magoo}.\foot{The importance
of these modes in AdS was independently recognized in \rasa, 
but they were not 
computed. They were computed in \mann, but only for a conformally invariant
scalar field whose asymptotic behavior is
similar to flat spacetime. The confining behavior of AdS is crucial for the
AdS/CFT correspondence.}
 According to the AdS/CFT
correspondence, a large static black hole in AdS corresponds to an
 (approximately)
thermal state in the CFT. Perturbing the black hole corresponds to perturbing
this thermal state, and the decay of the perturbation describes the
return to thermal equilibrium. So we obtain a prediction for the
thermalization timescale in the strongly coupled CFT. It seems difficult to 
compute this timescale directly in the CFT. Since the system will clearly
not thermalize in the free field limit, 
at weak coupling, this timescale will be very long and depend on the coupling
constant.
In the limit of strong coupling, it seems plausible that the
timescale will remain nonzero, and be independent of the coupling. This is
because the initial state is characterized by excitations with size of order
the thermal wavelength, so causality suggests that the relaxation timescale
should also be of order the thermal wavelength.

The results we obtain are consistent with this expectation. A black hole in AdS
is determined by two dimensionful parameters, the AdS radius $R$ and the black
hole radius $r_+$. The quasinormal frequencies must be functions of 
these parameters.
For large black holes, $r_+\gg R$, we will show that there is an
additional symmetry which insures that the frequencies can depend only on 
the black hole temperature $T \sim r_+/R^2$. However,
for smaller black holes, this
is no longer the case. Whereas the temperature begins to increase as one
decreases $r_+$ below $R$, we find that the (imaginary part of the) frequency
continues to decrease with $r_+$. This is different from what happens for
asymptotically flat black holes.
An ordinary
Schwarzschild black hole has only one dimensionful parameter which can
be taken to be the temperature. Its quasinormal frequencies must therefore
be multiples of this temperature. 
Thus small black holes in AdS do NOT
behave like black holes in asymptotically flat spacetime.
The reason is simply that
the boundary conditions at infinity are changed. More physically, the
late time behavior of the field is affected by waves bouncing off the
potential at large $r$.

Another difference from the asymptotically flat case concerns the decay
at very late times. For a Schwarzschild black hole,
it is known that the exponential
decay associated with the quasinormal modes eventually gives way to a 
power law tail \price. This has been shown to be associated with the
scattering of the field off the Coulomb potential at large $r$. As we will
discuss later, for
asymptotically AdS black holes, this does not occur.

We will compute the quasinormal frequencies for Schwarzschild-AdS black holes
in the dimensions of interest for the AdS/CFT correspondence: four, five, 
and seven. 
We will consider minimally coupled scalar perturbations representing,
e.g., the dilaton. This corresponds to a particular
perturbation of the CFT. For example,
for $AdS_5$, it corresponds to a perturbation of an (approximately) thermal
state in super Yang-Mills on $S^3 \times R$ with
$<F^2>$ nonzero. In the linearized
approximation we are using, the spacetime metric
is not affected by the scalar field. So the perturbation of the thermal
state does not change the energy density, which remains uniform over
the sphere.
The late time decay of this perturbation is universal in the sense that
all solutions for the dilaton with the same angular dependence will decay
at the same rate, which is  
determined by the imaginary part of the lowest quasinormal frequency.
Different perturbations, corresponding to different linearized
supergravity fields, will have different quasinormal frequencies  and
hence decay at different rates.
Although we work in the classical supergravity limit, our results would not
be affected if one includes small semiclassical corrections such as
black holes in equilibrium with their Hawking 
radiation.

A brief outline of this paper is the following.
In the next section we review the definition of quasinormal modes, their
relation to the late time behavior of the field, and derive
some of their properties using analytic arguments. The numerical approach
we use to compute the complex frequencies is described in section 3. In
the following section 
we discuss the results for both large black holes $r_+ \gg R$
and intermediate size black holes $r_+\sim R$. In section 5 we consider the
limit of small black holes $r_+ \ll R$. Although there is a striking similarity
between some of our results and some results obtained in the study of
black hole critical phenomena \chop,
we will argue that this is probably just a numerical
coincidence. The conclusion contains some speculations about the CFT 
interpretation of the quasinormal frequencies in the regime where they
do not scale with the temperature. In the appendix, we give 
some more details on our numerical 
calculations.

\newsec{Definition of quasinormal modes and analytic arguments}

Since we are interested in studying AdS black holes in various dimensions,
we begin with the $d$ dimensional Schwarzschild-AdS metric:
\eqn\sadss{ds^2 = - f(r) \, dt^2 + f(r)^{-1} dr^2 + r^2 \,  d\Omega_{d-2}^2 }
where
\eqn\f{
   f(r) \equiv  {r^2 \over R^2} +1 - \({r_0 \over r}\)^{d-3}. }
$R$ is the AdS radius and $r_0$ is related to the black hole mass
via 
\eqn\mass{M = {(d-2)A_{d-2}\ r_0^{d-3} \over 16\pi G_d}} 
where $A_{d-2} = 2 \pi^{d-1\over 2}/\Gamma({d-1\over 2})$ is the area
of a unit $(d-2)$-sphere.
The black hole horizon is at $r=r_+$, the largest zero of $f$,
and its Hawking temperature is
\eqn\hawtemp{T={f'(r_+)\over 4\pi} = {(d-1)r_+^2 + (d-3) R^2\over
 4\pi r_+ R^2}}

We are interested in solutions to the minimally coupled scalar wave equation
\eqn\wave{
   \nabla^2 \Phi = 0 }
If we consider modes 
\eqn\sepvar{
    \Phi(t,r,\rm{angles})
     =  r^{{2-d \over 2}} \, \psi(r) \, Y(\rm{angles}) \, 
          e^{-i \om t}} 
where $Y$ denotes the spherical harmonics on $S^{d-2}$,
and introduce a new radial coordinate
$dr_* = dr/f(r)$, the wave equation reduces to the standard  form  
\eqn\scat{[\p_{r_*}^2 +\om^2 -\tilde V(r_*)] \psi =0 .}
The potential $\tilde V$ is positive and vanishes at the horizon,
which corresponds to $r_*= -\infty$. It diverges at $r=\infty$,
which corresponds to a finite value of $r_*$.

To define quasinormal modes, let us first consider the case of a
simple Schwarzschild black hole.
Since the spacetime is asymptotically flat, the potential now vanishes
 near infinity. 
 Clearly, a solution exists
for each $\omega$ corresponding to a wave coming in from infinity, scattering
off the potential and being partly reflected and partly absorbed by the
black hole. Quasinormal modes are defined as solutions
which are purely outgoing near infinity $\Phi \sim e^{-i\omega (t-r_*)}$
and purely ingoing near the horizon $\Phi \sim e^{-i\omega (t+r_*)}$. No
initial incoming wave from infinity is allowed. This will
only be possible for a discrete set of 
complex $\omega$ called the quasinormal frequencies. 

For the asymptotically AdS
case, the potential diverges at infinity, so we must require that $\Phi$
vanish there.  In the absence of a black hole, $r_*$ has only a finite
range and solutions exist for only a discrete set of real $\omega$. However
once the black hole is added, there are again  solutions with any value
of $\omega$. These
correspond to an outgoing wave coming from the (past) horizon,
scattering off the potential and becoming an ingoing wave entering the
(future) horizon. Quasinormal modes are defined to be modes with only
ingoing waves near the horizon. These again exist for only a discrete
set of complex $\omega$. 

It should perhaps be emphasized that these modes are not the same as the
ones that have recently been computed in connection with the glueball masses
\glue. There are several differences: First, the background for
the glueball mass calculation is not the spherically symmetric AdS black 
hole, but an analytic continuation of the plane symmetric AdS black hole.
Second, because of the analytic continuation, the horizon becomes a regular
origin, and the boundary conditions there are not the
analytic continuation of the ingoing wave boundary condition imposed for
quasinormal modes. Finally, the glueball masses are real quantities, while
as we have said, the quasinormal frequencies will be complex. This makes them
more difficult to compute numerically.

One can show \refs{\clsy,\schmidt} that the complex \qn\
frequencies determine the fall off of the field at late times.
The basic idea is to start by writing the solution to the wave equation 
in terms of the retarded Green's function and initial data on a constant $t$
surface. One then rewrites the Green's function in terms of its Fourier
transform with respect to $t$. The quasinormal modes arise as poles
of the Green's function in the complex frequency plane, and their contributions
to the solution
can be extracted by closing the contour with a large semicircle  near infinity.

For a black hole in asymptotically flat spacetimes,
Price \price\ showed that
after the exponential decay due to the quasinormal ringing,
the field will decay as a power law $\Phi \sim t^{-(2l +3)}$ where
$l$ is the angular quantum number.
This has been seen explicitly in numerical simulations \gpp.
Mathematically, 
this is due to a cut in the Green's function along the negative imaginary
frequency axis. More physically,
this behavior
is due to scattering off the weak Coulomb potential near infinity.
For the case of a black hole in AdS, the potential diverges at infinity
and vanishes exponentially near the horizon.
Ching et. al. \clsy\ have analyzed the late time behavior of a broad class
of wave equations with potentials. They show that there are no power law
tails for a potential which vanishes exponentially. So there will be
no power law tails for black holes in AdS. 

For a black hole
with radius much smaller than the AdS radius, one might expect an
intermediate time regime where one sees power law behavior before the
new boundary conditions at infinity become important. However, this would
occur only if one starts with  large quasinormal modes with $\omega
\sim 1/r_+$ associated with a Schwarzschild black hole. We will see that
the lowest modes of a Schwarzschild-AdS black hole are much smaller and
their exponential decay is so slow that it eliminates 
the intermediate time power law behavior. 

The quasinormal frequencies will in general depend on the two parameters
in the problem $R,r_0$.  
By rescaling the metric, $\widehat{ds}^2 = \lambda^2 ds^2$,
and rescaling the coordinates $\hat t = \lambda t$ and
$\hat r = \lambda r$, the new metric again takes the form
\sadss\ with rescaled constants $R$ and $r_0$. Since the wave equation
\wave\ is clearly invariant under this constant rescaling of the metric,
we  can use it to set e.g. $R=1$. This rescaling is possible
for any metric and physically just corresponds to a choice of units.
In our case, we measure all
quantities in units of the AdS radius. The quasinormal frequencies can
still be arbitrary functions of $r_0$.

We now show that
for large black holes, $r_0 \gg R$, the frequencies must be proportional to 
the black hole temperature. This is a result of  an independent scaling
one can do in this limit. For large black holes,
the region outside the horizon of the
Schwarzschild-AdS metric \sadss\
becomes approximately plane symmetric:
\eqn\brane{ds^2 = - h(r) \, dt^2 + h(r)^{-1} dr^2 + r^2 \,  dx_i dx^i }
where
\eqn\h{
   h(r) \equiv  {r^2 \over R^2} - \({r_0 \over r}\)^{d-3}. }
For this metric  one can  rescale $r_0$  by a
pure coordinate transformation: $ t =  a\hat t,\  x_i =  a\hat x_i, \ 
 r =\hat r/ a$ for constant $a$.
 This does not rescale the overall metric, or the AdS radius
$R$.
The horizon radius $r_+^{d-1} = R^2 r_0^{d-3}$ gets rescaled by
$r_+ = \hat r_+/ a$. Of course, under this coordinate transformation
of the metric,
solutions
of the wave equation are related by the same coordinate transformation.
For solutions which are independent of $x^i$ (the analog of the $l =0$
modes) we have
$e^{-i\om(r_+) t} = e^{-i\om(\hat r_+) \hat t}$, which implies
$\om(r_+) \propto r_+$. Since the Hawking temperature of the metric \brane\
is also proportional to the horizon radius,
\eqn\temp{T={d-1\over 4\pi} {r_+\over R^2} }
we see that the frequencies
must scale with the temperature for large black holes. For solutions
proportional to $e^{ik_ix^i}$, this scaling argument implies
$\om(a r_+,ak_i) = a \om(r_+,k_i)$. So if $r_+^2 \gg k_ik^i$, one can 
rescale so that $k^2$ is negligibly small.
The above argument then shows that $\om$ still
scales with the temperature. One can then rescale back to $r_+ \gg R$ to apply
to large black holes. In other words, for any $k_i$, the quasinormal
frequencies scale with the temperature in the limit of large 
temperatures $T^2 \gg k^2$.
This argument
does not apply to black holes of order the AdS radius, and indeed we will find
that the quasinormal frequencies do not scale with the temperature in this
regime. But it does confirm the
expectation that the approach to thermal equilibrium in the dual
field theory should depend only on the temperature 
(at least for large temperature).  

Since we want modes which behave like
$e^{-i\omega (t+r_*)}$ near the horizon, it is convenient to set $v=t+r_*$,
and work with ingoing Eddington coordinates.
The metric for \SAdS in $d$ dimensions
 in ingoing Eddington coordinates is
\eqn\sads{
   ds^2 = - f(r) \, dv^2 + 2 \, dv \, dr + r^2 \,  d\Omega_{d-2}^2 }
where $f$ is again given by \f.
The minimally-coupled scalar wave equation \wave\
may be reduced to an ordinary, second order, linear 
differential equation in $r$ by the separation of variables,
\eqn\sepvar{
    \Phi(v,r,\rm{angles})
     =  r^{{2-d \over 2}} \, \psi(r) \, Y(\rm{angles}) \, 
          e^{-i \om v} 
}
This yields the following radial equation for $\psi(r)$:
\eqn\oder{
    f(r) \, {d^2 \over dr^2} \psi(r)
    + [f'(r) - 2 i \om]  \, {d \over dr} \psi(r) 
    - V(r)  \,  \psi(r) = 0, 
}
with the effective potential $V(r)$ given by ($R=1$)
$$V(r) = {(d-2)(d-4)\over 4r^2}f(r) + {d-2\over 2r} f'(r) + {c\over r^2} $$
\eqn\Veff{
     = {d (d-2) \over 4} + {(d-2)(d-4) + 4c \over 4 r^2} 
          + {(d-2)^2 r_0^{d-3} \over 4 r^{d-1}} }
where
\eqn\c{
    c = l (l + d - 3) }
is the eigenvalue of the Laplacian on $S^{d-2}$. Note that $V(r)$ is
manifestly positive for $d\ge 4$.
 
Ingoing modes near the (future) horizon are described, of course, by
a nonzero multiple of 
$e^{-i\om v}$. Outgoing modes  near the horizon
can also be expressed in terms of ingoing
Eddington coordinates via $ e^{-i\om (t-r_*)} = e^{-i\om v} e^{2i\om r_*}$.
Since 
\eqn\defrstar{ r_* = \int{dr\over f(r)}\approx {1\over f'(r_+)} \ln (r-r_+)}
near the horizon $r=r_+$, the outgoing
modes behave like
\eqn\outgoing{ e^{-i\om (t-r_*)} = e^{-i\om v} e^{2i\om r_*} 
  \approx e^{-i\om v} (r-r_+)^{2i\om/f'(r_+)}}
Since $v,r$ are good coordinates near the horizon,
the outgoing modes are not smooth ($C^\infty$) at $r=r_+$
unless $2i\om/f'(r_+)$ is a positive integer.
We show below that the imaginary part of $\om$ must be negative, so the 
exponent in \outgoing\ always has a positive real part. Thus the outgoing
modes vanish near the future horizon, while the ingoing modes are nonzero
there. However we also show (in the next section) that $2i\om/f'(r_+)$ cannot be
a positive integer, so the outgoing modes are not smooth at $r=r_+$.

We wish to find the complex values of $\om$ such that \oder\ has a solution
with only ingoing modes near the horizon, and vanishing at infinity. 
We will eliminate the outgoing modes by first
assuming the solution is smooth at  $r=r_+$, and then
showing that the
allowed descrete values of $\om$ are such that $2i\om/f'(r_+)$
is not an integer. The actual values of $\om$ must be computed 
numerically, but some general properties can be seen
analytically. For example, we now show that
there are no solutions with $i\omega$ pure 
real, and $2i\omega < f'(r_+)$. If $i\omega$ were  real, then the
equation would be real and the solutions $\psi$  would be real. If there were
a local extremum at some point $\tilde r$, then $\psi'(\tilde r)=0$ and 
$\psi''(\tilde r)$ would have the same sign as $\psi(\tilde r)$.
So if $\psi$ were 
 positive at $\tilde r$, it would have to increase as $r$ increased. Similarly,
if it were negative, it would have to decrease. In neither case, could it
approach zero asymptotically. We conclude that the solutions 
must monotonically approach zero.  Now if $2i\omega < f'(r_+)$,
$\psi'(r_+)$ has the same sign as  $\psi(r_+)$.\foot{This is where the
condition of no outgoing modes near the horizon is used. If outgoing waves
were present, $f(r) \psi''(r)$ would no longer vanish at $r=r_+$, and
$\psi'(r_+)$ need not have the same sign as $\psi(r_+)$.}
So as one moves away from the 
horizon, the solutions move farther away from
zero and hence can never reach zero asymptotically. 
This analytic argument only applies if $2i\omega < f'(r_+)$. But we will
see numerically that even without this restriction, there are no solutions
with $i\om$ pure real.

A more powerful result can be obtained
as follows. Multiplying \oder\ by $\bar\psi$ and integrating 
from $r_+$ to $\infty$ yields
\eqn\iden{ \int_{r_+}^\infty dr \[\bar\psi {d \over dr}\( f {d\psi \over dr}\)
  -2i\om \bar\psi {d\psi \over dr} - V \bar\psi \psi \] =0 }
The first term can be integrated by parts without picking up a surface term
since $f(r_+)=0$ and $\bar\psi(\infty)=0$. This yields
\eqn\xx{\int_{r_+}^\infty dr [f|\psi'|^2 +2i\om \bar\psi \psi' + V |\psi|^2] =0}
Taking the imaginary part of \xx\ yields
\eqn\imag{ \int_{r_+}^\infty dr [\om \bar\psi \psi' + \bar \om \psi \bar \psi'] 
=0}
Integrating the second term by parts yields
\eqn\almost{ (\om - \bar \om) \int_{r_+}^\infty dr \bar\psi \psi' =  \bar \om
|\psi(r_+)|^2  }
Substituting this back into \xx\ we obtain the final result
\eqn\final{ \int_{r_+}^\infty dr [f|\psi'|^2 + V |\psi|^2 ] = 
-{|\om|^2 |\psi(r_+)|^2 \over {\rm Im} \om}  }
Since $f$ and $V$ are both positive definite outside the horizon,
this equation clearly shows that there are no solutions with Im $\om >0$.
These would correspond to unstable modes which grow exponentially in time.
There are also no solutions with Im $\om =0$: All solutions must decay in time.
In addition, eq. \final\ shows that the only solution which vanishes at the 
horizon (and infinity) is zero everywhere.
Since the equation is linear, we can always
rescale $\psi$ so that $\psi(r_+)=1$.

\newsec{Numerical approach to computing quasinormal modes}

To compute the quasinormal modes, we will expand the solution in a power
series about the horizon and impose the boundary condition that
the solution vanish at infinity. In order to map the entire region of interest,
$r_+ < r < \infty$, into a finite parameter range, we change variables
to $x = 1/r$. In general, a power series expansion  will
have a radius of convergence at least as large as the
distance to the nearest pole.
Examining the pole structure of \oder\ 
in the whole complex $r$ plane, 
we find  $d+1$ regular singular points, at
$r=0$, $r=\infty$, 
and at the $d-1$ zeros of $f$, 
one of which, $r = r_+$ corresponds to the horizon.
At least for $d=4,5$ or $7$, if we use the variable $x = 1/r$ and
expand about the horizon, $x_+ = 1/r_+$,
the radius of
convergence will\foot{For $d=4$ and $d=5$, one can show analytically
that starting at the horizon, $x=x_+$, the nearest pole is indeed $x=0$.
For $d=7$ we have checked numerically that this is again the case.}
reach to $x=0$, so that we can use this expansion to 
consider the behavior of the solution as $r \to \infty$.

In terms of our new variable $x = 1/r$, \oder\ becomes
\eqn\odex{
    s(x) \, {d^2 \over dx^2} \psi(x)
    + {t(x) \over x-x_+}  \, {d \over dx} \psi(x) 
    + {u(x) \over (x-x_+)^2}  \, \psi(x) =0 
}
where the coefficient functions are given by
\eqn\s{
    s(x) = { r_0^{d-3} \,  x^{d+1} - x^4 - x^2 \over  x-x_+ } 
     = {x_+^{2} + 1 \over x_+^{d-1}} x^{d} +\ldots + 
       {x_+^{2} + 1 \over x_+^{3}} x^{4} + {1 \over x_+^{2}} x^{3}
       + {1 \over x_+} x^{2}
}
\eqn\t{
    t(x) = (d-1) \,  r_0^{d-3} \,  x^{d} - 2x^3  - 2x^2 \, i \om
}
\eqn\u{
    u(x) = (x-x_+) \, V(x)
}
The parameter $r_0^{d-3}$ should be viewed as a function of the horizon
radius: $r_0^{d-3} = {x_+^{2} + 1\over x_+^{d-1}}$.
Since $s$, $t$, and $u$ are all polynomials of degree $d$
we may expand them about the horizon $x=x_+$:
$s(x) = \sum_{n=0}^d s_n \, (x - x_+)^n$,
and similarly for $t(x)$ and $u(x)$.
It will be useful to note that 
$s_0 = 2 x_+^2 \, \kappa$, 
$t_0 = 2 x_+^2 \, (\kappa - i \om)$, and 
$u_0 = 0$,
where $\kappa$ is the surface gravity, which is related to the black hole
temperature \hawtemp\ by 
\eqn\surgrav{\kappa ={f'(r_+)\over 2} = 2 \pi T.} 
Also, since $s_0 \ne 0$,
$x = x_+$ is a regular singular point of \odex.

To determine the behavior of the solutions near the horizon, 
we first set $\psi(x) = (x-x_+)^\alpha$ and substitute into \odex.
Then to leading order we get
\eqn\indicial{    \alpha (\alpha -1)  \, s_0 + \alpha \, t_0 =
    2 x_{+}^2 \alpha \, \( \alpha \, \kappa -i \om \) =0}
which has two solutions $\alpha=0$ and $\alpha = i\om/\kappa$. 
We see from \outgoing\ that these correspond
precisely to the ingoing and outgoing modes near the horizon respectively.
Since we want to include only the ingoing modes, we take $\alpha=0$.
This corresponds to 
looking for a solution of the form
\eqn\psix{
    \psi(x) =  
    \sum_{n=0}^{\infty} a_n \, (x - x_+)^n  
}
Substituting \psix\ into \odex\ and equating coefficients of 
$(x - x_+)^n$ for each $n$,
 we obtain the following recursion relations\foot{
Although the standard way of writing \odex\ is to set the coefficient
of $\psi''$ to $1$ which yields simpler-looking recursion relations, 
the advantage of the present formulation is that
since $s(x)$, $t(x)$, and $u(x)$ are polynomials, their analytic 
expansions will terminate after a finite number of terms, so that
each $a_n$ will be given in terms of a relatively small number of terms.
}
 for the $a_n$:
\eqn\recurs{
    a_n = -{1 \over P_n} \,  \sum_{k=0}^{n-1} 
    \[ k (k -1)  \, s_{n-k}
    + k \, t_{n-k} + u_{n-k} \] \, a_k
}
where 
\eqn\Pn{
    P_n = 
    n (n-1)  \, s_0 + n \, t_0 =
    2 x_{+}^2 n \, \( n \, \kappa -i \om \)
}
Since the leading coefficient $a_0$ is undetermined, this yields a 
one parameter family of solutions, as expected for a linear equation.

The solutions to \wave\ in asymptotically AdS spacetime are
$\Phi \sim \rm{constant}$ and $\Phi \sim 1/ r^{d-1}$ as $r \to \infty$, 
which translates into 
$\psi \sim  r^{d-2 \over 2}$ and $\psi \sim r^{-d/2}$, respectively.
We are interested in normalizable modes, so 
we must select only solutions which 
satisfy $\psi \to 0$ as $r \to \infty$ (or $x \to 0$).
This means that we require \psix\ to vanish at $x = 0$,
which is satisfied only for special (discrete) values of $\om$.
(For all other values of $\om$, the solution will blow up,
$\psi(0)  = \infty$.)
Thus in order to find the \qn\ modes, we need to find
the zeros of $\sum_{n=0}^{\infty} a_n( \om) \, (- x_+)^n$ 
in the complex $\om$ plane. This is done by truncating the series after
a large number of terms  and computing the partial sum as a function of
$\om$. One can then find zeros of this partial sum, and check the accuracy
by seeing how much the location of the zero changes as one goes to higher
partial sums. Some details are given in the Appendix.

One can now easily show that 
$2i\om/f'(r_+) = i\om/\kappa$
cannot be an integer.
 If $\om$ is pure imaginary and $i\om = \tilde n\kappa$ for some integer
 $\tilde n$,
then $P_{\tilde n} =0$.
This implies an additional constraint on the coefficients
$a_k$, $k = 0, \cdots, \tilde n-1$ which will only be satisfied if
they vanish. In other words, the solution will behave like $(x-x_+)^{\tilde n}$
near the horizon corresponding to a pure outgoing wave. However, since $\psi$
now vanishes at the horizon,
\final\ implies that $\psi$ vanishes everywhere. So there are no nontrivial
solutions with $i\om /\kappa$ equal to an integer. As we saw in section two,
this means that if
one wanted to include outgoing modes near the (future) horizon, the solution
would not
be smooth there.

\newsec{Discussion of results}

The numerical procedure described above 
can be applied to both
large black holes ($r_+ \gg R$) and intermediate size
black holes ($r_+ \sim R$). In this section we describe the results.
We set $R =1$, and decompose the \qn\ frequencies into real and 
imaginary parts:
\eqn\defRI{ \om = \om_R - i \om_I}
With the sign chosen in \defRI, $\om_I$ is positive for all \qn\ frequencies.
%
%
\midinsert
\centerline{%
\vbox{
  \offinterlineskip \tabskip=0pt
  \halign{\strut
          \vrule#&              %
          \hfil $ #~$ &\vrule#& %
          \hfil $\,#$ &         %
        ~ \hfil $#$ &\vrule#&   %
          \hfil $\,#$ &         %
        ~ \hfil $#$ &\vrule#&   %
          \hfil $\,#$ &         %
        ~ \hfil $#$ &\vrule#&   %
          \hfil $\,#$ &         %
        ~ \hfil $#$ &\vrule#    %
          \cr
     \noalign{\hrule}
     & \hfill &&  \multispan 2 \hfil  4d BH modes  \hfil &&
                  \multispan 2 \hfil  5d BH modes  \hfil &&
                  \multispan 2 \hfil  7d BH modes  \hfil  &     \cr
     \noalign{\hrule}
 & \omit ~$r_+$
       &&  \omit \hfil $\om_I$ \hfil & \omit \hfil $\om_R$ \hfil && 
           \omit \hfil $\om_I$ \hfil & \omit \hfil $\om_R$ \hfil &&
           \omit \hfil $\om_I$ \hfil & \omit \hfil $\om_R$ \hfil & \cr
     \noalign{\hrule}
    &   100    &&  266.3856     &  184.9534   && 274.6655   &  311.9627  && 261.2    & 500.8   &                \cr
    &    50    &&  133.1933     &   92.4937   && 137.3296   &  156.0077  && 130.7    & 250.4   &                \cr
    &    10    &&   26.6418     &   18.6070   &&  27.4457   &  31.3699   &&  26.07   & 50.35   &                \cr
    &     5    &&   13.3255     &    9.4711   &&  13.6914   &  15.9454   &&  12.96  & 25.57  &                \cr
    &     1    &&    2.6712     &    2.7982   &&   2.5547   &   4.5788   &&  2.16   &  7.27    &                \cr
    &     0.8  &&    2.1304     &    2.5878   &&   1.9676   &   4.1951   &&          &         &                \cr
    &     0.6  &&    1.5797     &    2.4316   &&   1.3656   &   3.8914   &&          &         &                \cr
    &     0.4  &&    1.0064     &    2.3629   &&   0.7462   &   3.7174   &&          &         &                \cr
     \noalign{\hrule}
                                                             }}}
\smallskip
Table 1: The lowest \qn\ mode frequency for the 4, 5, and 7 dimensional
\centerline{\SAdS\ black hole for some selected black hole sizes.} 
\endinsert

\ifig\ldlg{For large black holes, $\om_I$ is proportional to the temperature.
The top line is $d=4$, the middle line is $d=5$ and the bottom line is 
$d=7$.}
{\epsfxsize=9.5cm \epsfysize=5.5cm \epsfbox{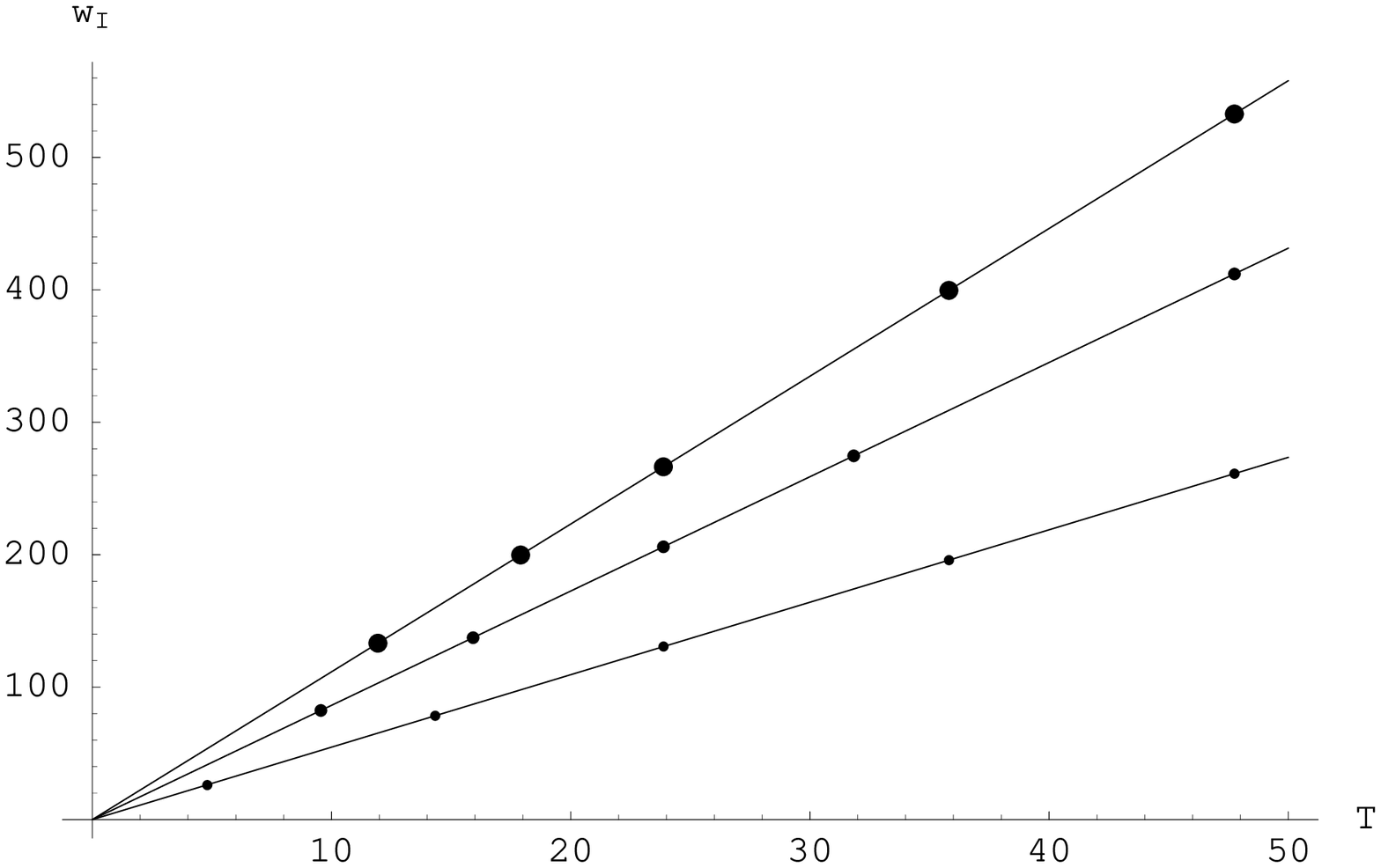}}

\ifig\lolg{For large black holes, $\om_R$ is also
proportional to the temperature.
The top line is now $d=7$, the middle line is $d=5$ and the bottom line is
$d=4$.}
{\epsfxsize=9.5cm \epsfysize=5.5cm \epsfbox{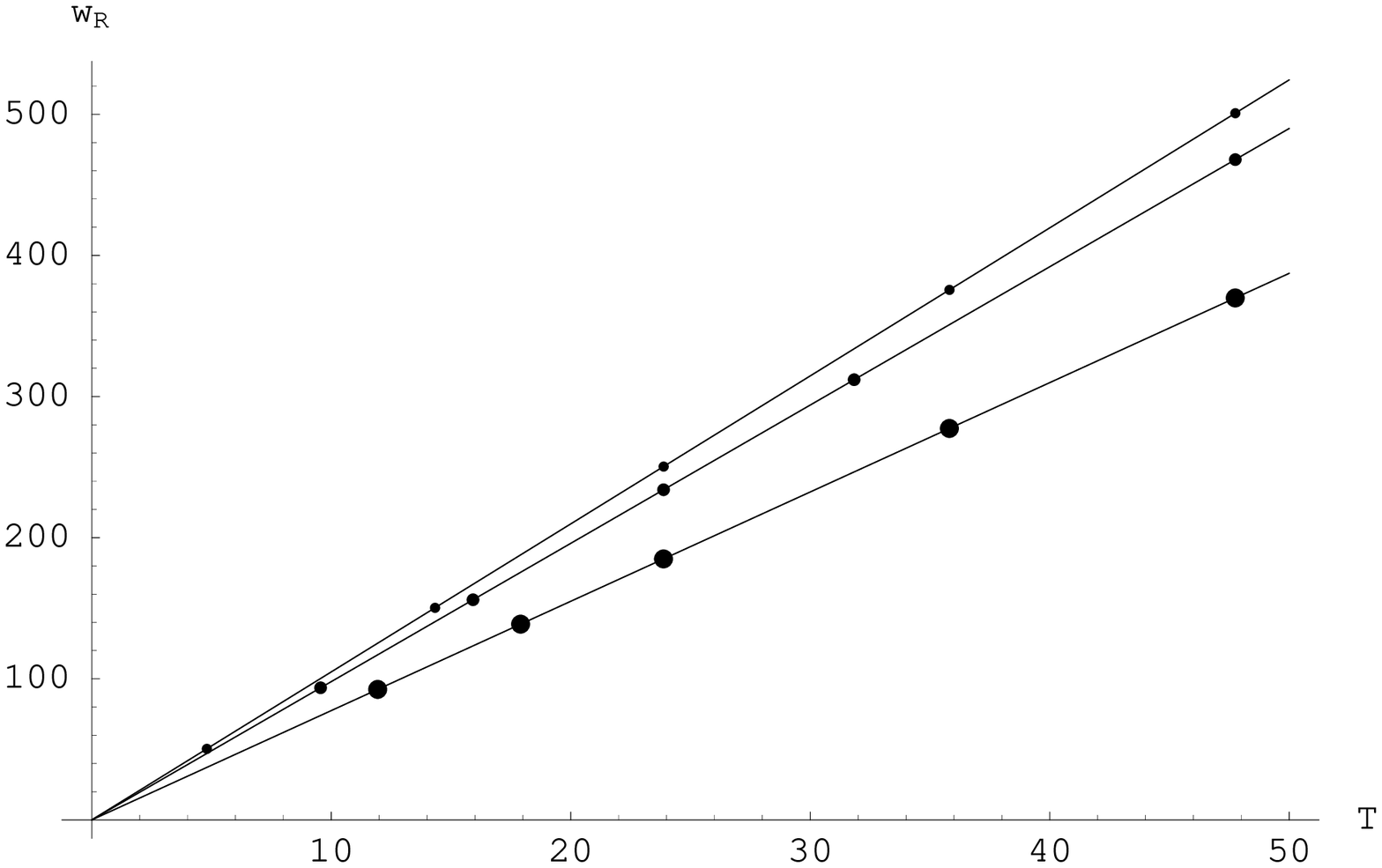}}

In Table 1., we list the values of the lowest \qn\ mode frequencies for
$l=0$ and selected
values of $r_+$, for the four, five, and seven dimensional \SAdS black holes.
{} For large black holes,
both the real and the imaginary parts of the frequency 
are linear functions of $r_+$. Since the temperature of a large black hole 
is $T= (d-1)r_+/4\pi$, it follows that they are also linear functions of $T$.
This is clearly  shown in \ldlg\ and \lolg, where $\om_I$ and $\om_R$
respectively
are plotted as a function of the temperature
for the four, five, and seven dimensional cases. 
The dots, representing the 
\qn\ modes, lie on straight lines through the origin.
 In \ldlg, the top line corresponds to the $d=4$ case, 
the middle line is the $d=5$ case, 
and the bottom line is the  $d=7$ case.
Explicitly, the lines are given by
$$\om_I = 11.16 \ T \qquad {\rm for} \ d=4$$
$$\om_I = 8.63 \ T \qquad {\rm for} \ d=5$$
\eqn\qntemp{\om_I = 5.47 \ T \qquad {\rm for} \ d=7}
Notice from Table 1 that as a function of $r_+$, $\om_I$ is almost independent
of dimension. The difference in these slopes is almost entirely due
to the dimension dependence of the relation between $r_+$ and $T$ \temp.
In contrast, $\om_R$ does depend on the dimension, and in 
\lolg, the order of the lines is reversed: 
$$\om_R = 10.5 \ T \qquad {\rm for} \ d=7$$
$$\om_R = 9.8 \ T \qquad {\rm for} \ d=5$$
\eqn\qnrtemp{\om_R = 7.75 \ T \qquad {\rm for} \ d=4}
This linear
scaling with the temperature is in agreement with the general argument in 
section 2. According to the AdS/CFT correspondence,  $\tau=1/\om_I$ is the
timescale for the approach to thermal equilibrium. Eq. \qntemp\ is one of
the main results of this work.

\ifig\fsdiv{$\om_I$ for intermediate black holes in four dimensions.
The solid line is 
$\om_I = 2.66\ r_+$, and the dashed line is $\om_I = 11.16\ T$.}
{\epsfxsize=9.5cm \epsfysize=5.5cm \epsfbox{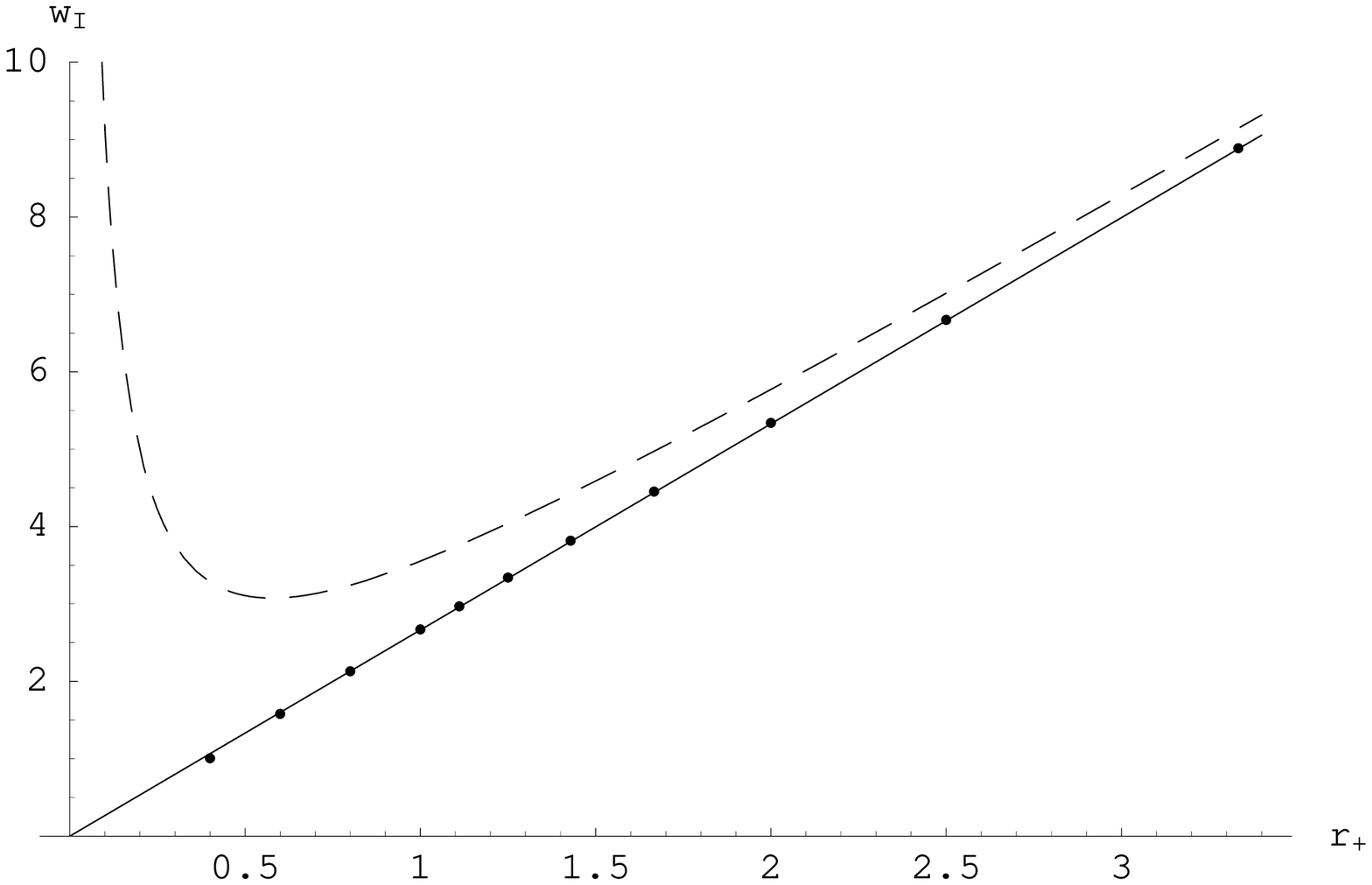}}

\ifig\fsdv{$\om_I$ for intermediate black holes in five dimensions.
The solid line is
$\om_I = 2.75\ r_+$, and the dashed line is $\om_I = 8.63\ T$.}
{\epsfxsize=9.5cm \epsfysize=5.5cm \epsfbox{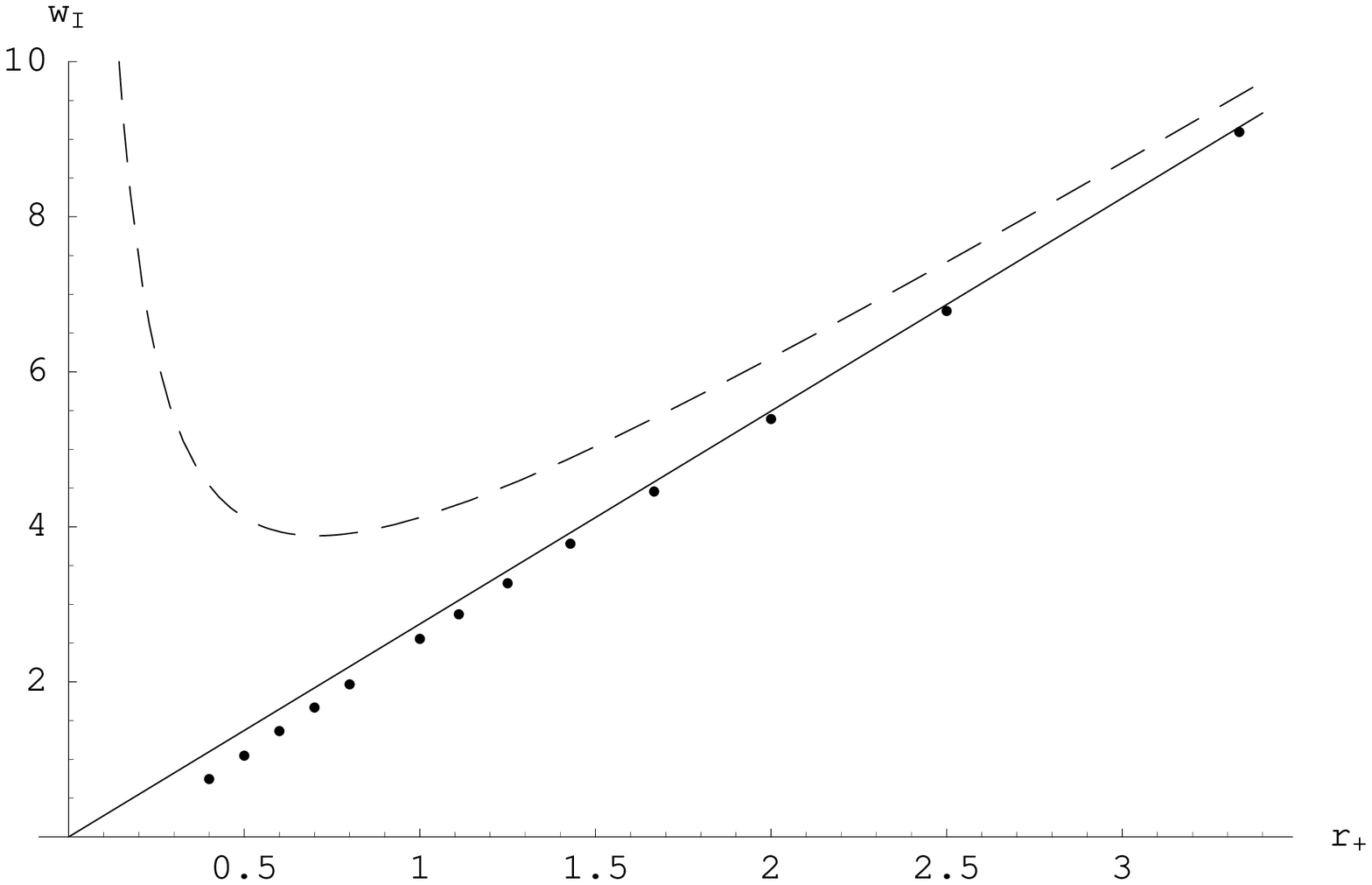}}

For the intermediate size 
black holes, 
the \qn\ frequencies do not scale with the temperature.
This is clearly shown in \fsdiv\ which plots $\om_I$ as a function 
of $r_+$ for $d=4$ black holes with $r_+ \sim 1$. To a remarkable
acuracy, the points continue to lie along a  straight line
$\om_I = 2.66\ r_+$. The dashed curve represents the continuation of
the curve $\om_I = 11.16\ T$ shown in fig. 1. to smaller values of 
$r_+$. (For large $r_+$ these two
curves are identical.) It is not yet clear what the significance of this linear
relation is for the dual CFT. Some speculations are given in section 6.
Since the \qn\ frequencies can be computed to an accuracy
much better than the size of the dots in \fsdiv, one can check that 
the points
actually lie slightly off the line. This is  shown more clearly in 
the five dimensional results in \fsdv. Once again the dashed curve
is the continuation of the curve $\om_I = 8.63\ T$ shown in fig. 1,
and the solid curve is the line  $\om_I = 2.75\ r_+$ that it approaches
asymptotically.

\ifig\fsoiv{$\om_R$ for intermediate black holes in four dimensions.
The solid line is 
$\om_R = 1.85\ r_+$, and the dashed line is $\om_R = 7.75\ T$}
{\epsfxsize=9.5cm \epsfysize=5.5cm \epsfbox{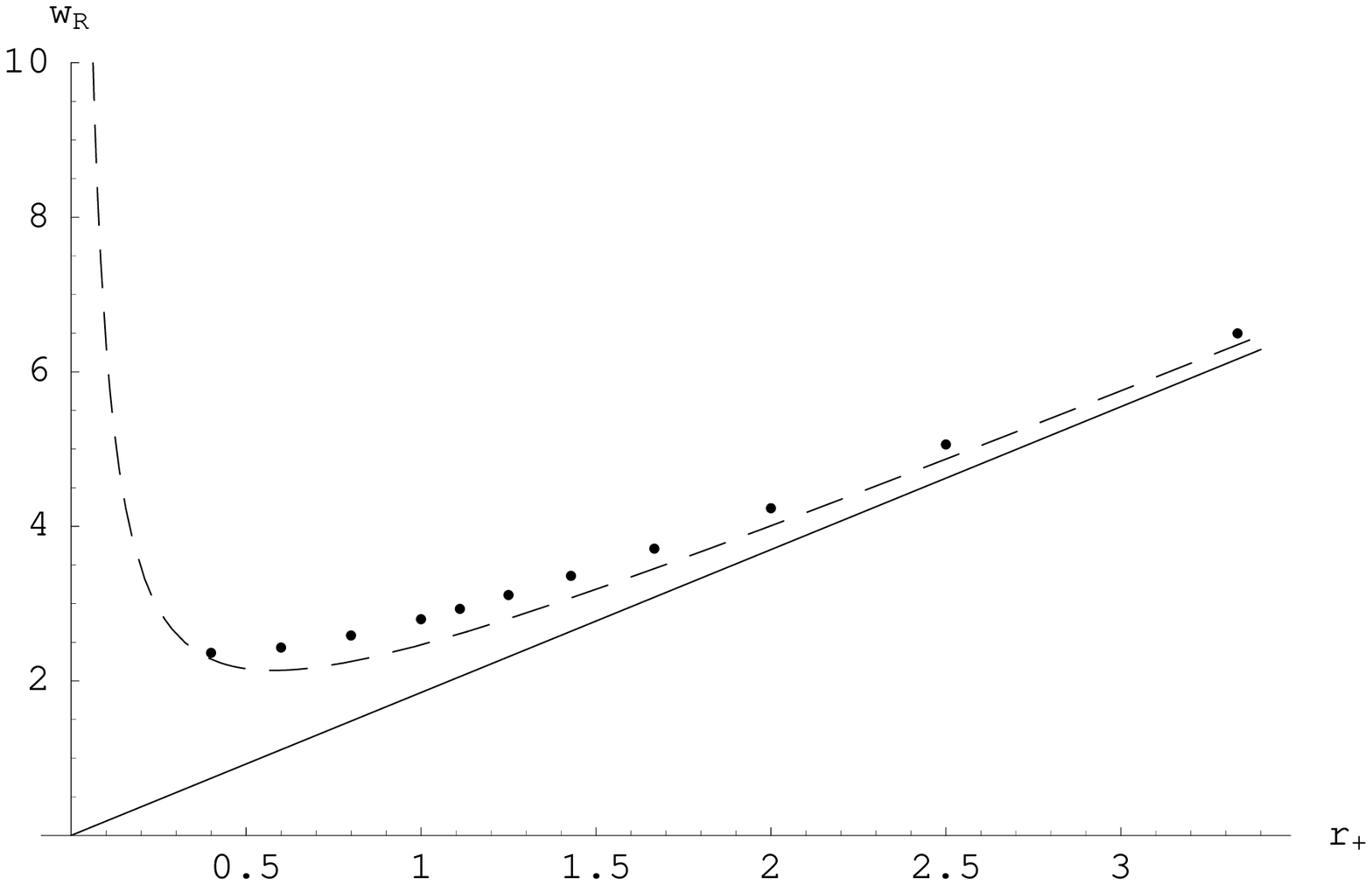}}

\ifig\fsov{$\om_R$ for intermediate  black holes in five dimensions.
The solid line is
$\om_R = 3.12\ r_+$, and the dashed line is $\om_R = 9.8\ T$}
{\epsfxsize=9.5cm \epsfysize=5.5cm \epsfbox{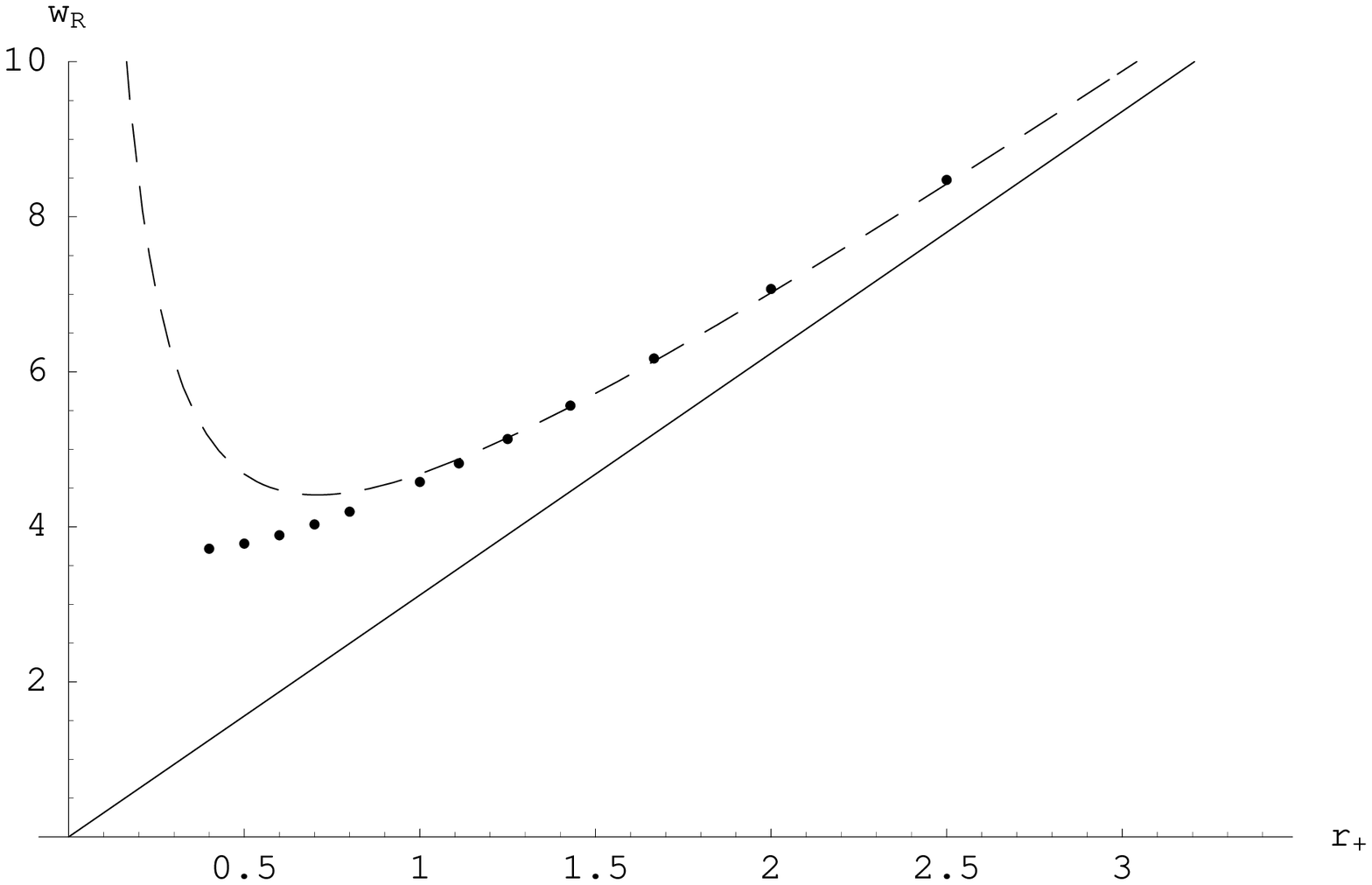}}

The real part of the \qn\ frequencies are shown in similar plots
in \fsoiv\ for $d=4$ and \fsov\ for $d=5$.
$\om_R$ approximates the temperature
more closely than the black hole size, but it is clear from \fsov\
that it is not diverging for small black holes.

We have so far discussed only the lowest \qn\ mode with $l=0$. We have 
also computed higher modes and modes with nonzero angular momentum, but
the numerical accuracy decreases as one increases the mode number $n$ or
$l$. So we restrict our attention to relatively small values of $n$ and $l$.  

For large black holes, in both four and  five dimensions,  
we find that the low lying \qn\ modes are approximately evenly spaced in $n$.
In particular, for $r_+=100$, $\om_I(n) \approx 41 +  225 \, n$ and 
 $\om_R(n) \approx 54 + 131 \, n$ in four dimensions, whereas
$\om_I(n) \approx 73 +  201 \, n$ and 
 $\om_R(n) \approx 106 + 202 \, n$ in five dimensions.

\ifig\flvar{Dependence of $\om$ on $l$ for four dimensional
black hole with $r_+ =1$.
The smaller points are $\om_R$, the larger points are $\om_I$.}
{\epsfxsize=9.5cm \epsfysize=5.5cm \epsfbox{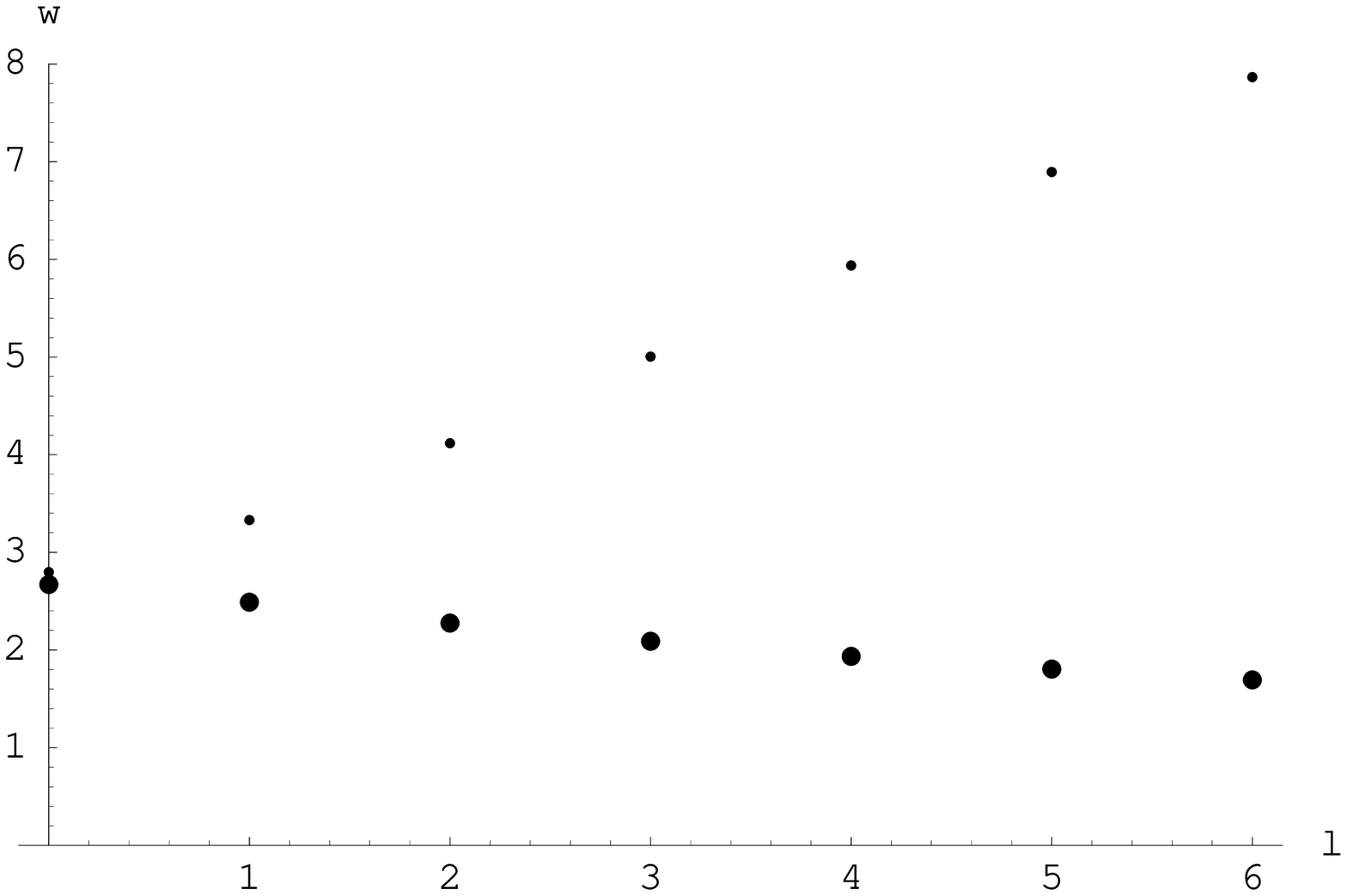}}

Increasing the angular momentum $l$ mode has the surprising 
effect of increasing the damping time scale (\ $\om_I$ decreases),
and decreasing the oscillation time scale ($\om_R$ increases).
This is shown in \flvar, where $\om_R$ (smaller points)
and $\om_I$ (larger points) are plotted against $l$ for
low values of $l$.\foot{The size of the dots is not related to the
accuracy of the calculation.}
An important open question is the behavior of $\om_I$ as $l\rightarrow
\infty$. It appears to decrease with $l$, but the general argument
in section 2 shows that it cannot become
negative. If $\om_I$ continues to decrease with $l$, then the late time
behavior of a general perturbation will be dominated by the largest $l$
mode. The large $l$ behavior of $\om_I$ is currently under investigation.
Preliminary results indicate that the frequencies stay bounded away from zero.
If this were not the case, and $\om_I$ approached zero fast enough, 
then 
a general superposition of all spherical harmonics could decay at late times
only as a power law.  However, even this would not be a problem for the AdS/CFT
correspondence, since the  decomposition into spherical harmonics can be done
in the boundary 
field theory as well. The statement is that, e.g., a perturbation 
of $<F^2>$ with given angular dependence $Y_l$ on $S^3$ will decay exponentially
with a time scale given by the imaginary part of the lowest quasinormal
mode with that value of $l$.

\newsec{Comments on small black holes}

In this section we briefly discuss the extrapolation of the \qn\
frequencies to the small black hole regime ($r_+\ll R$). Our numerical
approach becomes unreliable in this regime, so we cannot compute them
directly. Instead, we must rely on indirect arguments. But first we give
some motivation for exploring this question.

Small AdS black holes are not of direct interest for
the AdS/CFT correspondence. This is because 
an extended black hole of the form
\SAdS\ cross $S^m$ is unstable to forming
a black hole localized in all directions whenever the radius of the
black hole is smaller than the radius of the sphere. This is a classical
instability first discussed by Gregory and Laflamme \grla.
It is quite different
from the Hawking-Page transition \refs{\hapa,\witt}
which applies to black holes in contact with
a heat bath. In that case, when the black hole is of order the AdS radius
it undergoes a transition to a thermal gas in AdS. The Hawking-Page 
transition can be avoided if we consider states of fixed energy,
not fixed temperature.
Then black holes dominate the entropy even
when $r_+ <R$ and continue to do so until $r_+/R$ is less than a negative power
of $N$ \refs{\bdhm,\gary}. 
The situation is very similar to the old studies of a black hole
in a box. For fixed total energy, the maximum entropy state consists of
most of the energy in the black hole, and a small amount in  radiation.
Unfortunately, the stable small black hole configuration must
be a ten or eleven dimensional black hole (by the Gregory-Laflamme instability)
and is not known explicitly.

Nevertheless, there may be other applications of the quasinormal modes
of small black holes in AdS. One possibility comes from the striking
fact (shown in \fsdiv)   that for $d=4$,  $\om_I$ is proportional
to $r_+$ to high accuracy. As we will discuss below, 
the slope of this line, $2.66$, turns out to be numerically very
close to a special frequency which arises in black hole critical phenomena
first studied by Choptuik \chop.
To explore this possible connection, one needs to
consider \qn\ modes of small black holes.

From the intermediate black hole results shown in the previous section,
it is tempting to speculate that as $r_+\rightarrow 0$, $\om_I
\rightarrow 0$, and $\om_R\rightarrow $ constant. Since the decay of
the field is due to absorption by the black hole, it is intuitively plausible
that as the black hole becomes arbitrarily small,
the field will no longer decay.
It is even possible that the \qn\ modes approach the usual 
AdS modes in the limit $r_+\rightarrow 0$, although this is not guaranteed since
the boundary conditions at $r=r_+$ do not reduce to regularity at the origin
as $r_+\rightarrow 0$. If they do approach  the usual modes in this limit,
then $\om_R$ must approach  $d-1$ \bulu. Of course, in the context of
string theory, one cannot trust the Schwarzschild-AdS solution when the
curvature at the horizon becomes larger than the string scale. By taking the
AdS radius $R$ sufficiently large, one can certainly use this solution to
describe some small black holes,  but the geometry
would have to be modified before the limit $r_+\rightarrow 0$ is reached. 

It has been shown that the low energy absorption cross section 
for massless scalars incident on a general asymptotically flat
spherically symmetric black hole
is always equal to the area of the event horizon \dgm. We can use this to 
estimate the imaginary part of the lowest quasinormal mode for a small AdS
black hole as follows. Imagine a wave with energy of order $1/R$ propagating
toward a black hole with $r_+ \ll R$. Then the spacetime around
the black hole is approximately Schwarzschild, and the low energy
condition is 
satisfied, so the  amplitude of the reflected wave $\Phi_r$  will be reduced
from the amplitude of the incident wave $\Phi_i$, by $1-(\Phi_r/\Phi_i)^2 
\sim r_+^{d-2}$. After a time of order the AdS radius, the reflected wave 
will bounce off the potential at infinity with no change in amplitude.
It will again encounter the black hole potential and be partly absorbed 
and partly reflected. Repeating this process leads to a gradual decay of 
the field $\Phi \sim e^{- \alpha  t}$ with $\alpha \sim r_+^{d-2}$.

This suggests that for small black holes, $\om_I$ should scale like
the horizon area $r_+^{d-2}$.
In the large black hole regime, on the other hand, we know that 
the modes should scale linearly with $r_+$. To check this,
we consider a simple ansantz which interpolates between these two regimes
and see how well it fits the data. Consider the function $\om_I(r_+) = 
{a r_+^m \over b + r_+^{m-1} }$
(where $a$ corresponds to the asymptotic slope). For each $m$
we choose $b$ to give the best fit to the intermediate black hole data,
and see which $m$ yields the lowest overall error (as measured by $\chi^2$).

\ifig\ffitv{The curved line is a fit to the modes of a small black hole in 
$d=5$. The modes approach the straight line shown at large $r_+$.}
{\epsfxsize=9.5cm \epsfysize=5.5cm \epsfbox{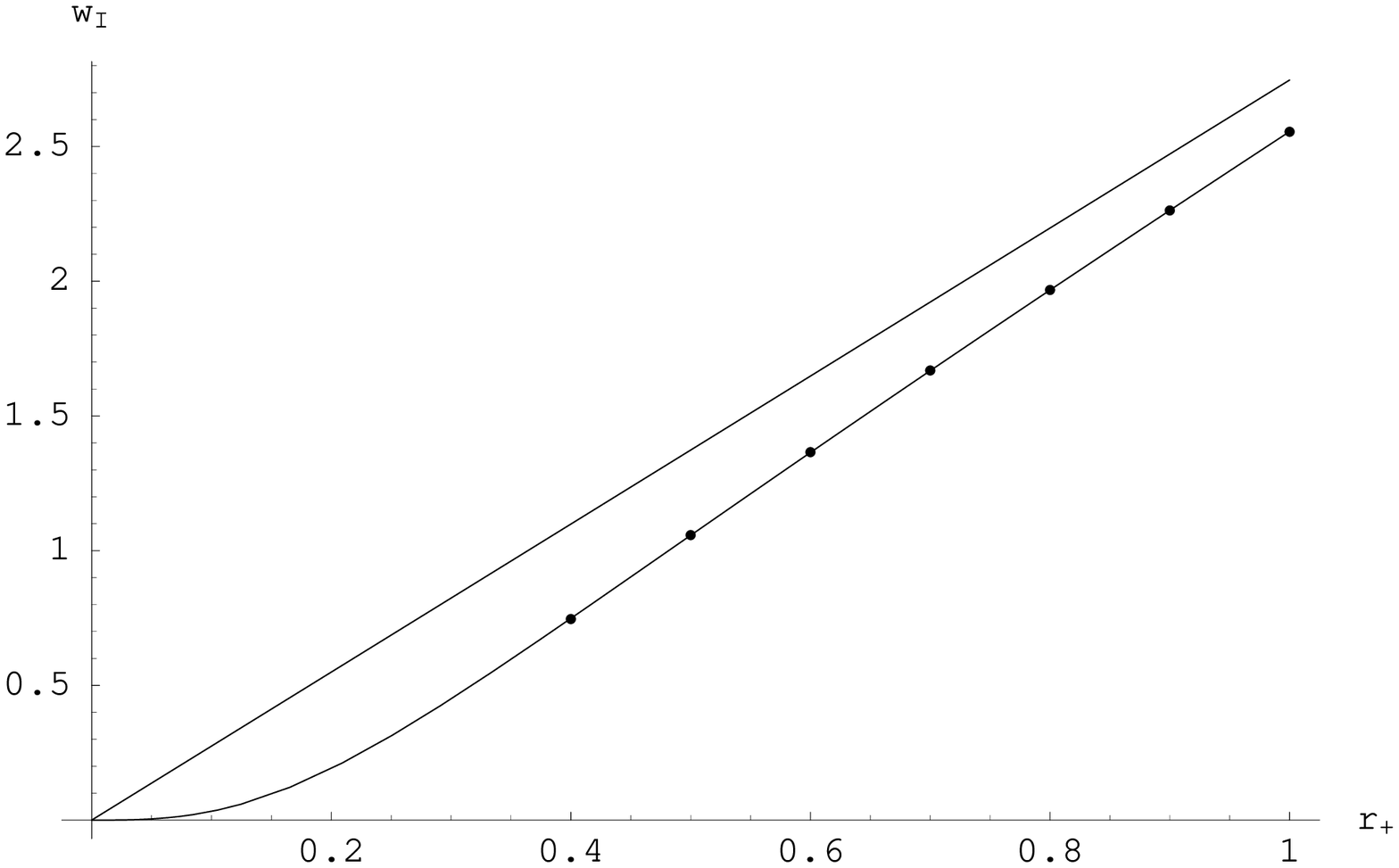}}

In the five dimensional case, using seven points between $r_+ =.4$ and
$r_+ =1$, we indeed find that $m=3$ gives the best fit:
 $\chi^2 \approx 9 \times 10^{-6}$ for $m=3$,
as opposed to $\chi^2 \approx 3 \times 10^{-2}$ for $m=2$ and $m=4$.
The actual fit, shown in \ffitv\ along with the  modes and
the asymptotic line, is given by 
$\om_I(r_+) \approx {2.746675 r_+^3 \over 0.0748 + r_+^2 }$.
In four dimensions the story is much less clear, since there is no
significant difference between the fit with $m=2$ and the fit with
$m=3$. This could be due to the fact  that the data for intermediate
black holes have not yet started to significantly deviate from
a straight line.

 To see the possible connection with black hole critical phenomena,
consider the evolution of a self gravitating spherically symmetric
scalar field (in an asymptotically flat 4d spacetime). 
It is clear that weak waves will scatter and go off to infinity,
just like in
flat spacetime. Strong waves will collapse and form a black hole.
Choptuik \chop\ studied 
one parameter families of initial data (labelled by $p$)
which interpolated between these two extremes. In each case there is a critical
solution $p=p_*$
which marks the boundary between forming a black hole  or not forming
one. The late time behavior of this critical solution turns out to be universal.
It has precisely one unstable mode which grows like $e^{\lambda t}$ with
$\lambda = 2.67$. This mode is responsible for
the famous scaling of the black hole mass for $p$ just above the critical
value, $M_{bh} \sim (p-p_*)^\gamma$ where $\gamma = 1/\lambda = .374$.
(For a review, see \gund.)

The numerical value of $\lambda$ is very close to the slope, $2.66$, of 
the line in \fsdiv\  giving the imaginary part of the quasinormal mode
frequencies.
Since both numbers involve imaginary frequencies
for  spherically symmetric scalar
fields in four dimensions, it is natural to wonder if there might be
a deeper connection between these two phenomena. 
Unfortunately, it appears at the moment that the agreement is just
a numerical coincidence.
The first thing one might check is whether
the agreement continues in higher dimensions.
Although the critical solutions for black hole formation
have not been studied in five or seven dimensions, they have recently been
calculated in six dimensions \gcd\ with the result $\lambda =1/.424= 2.36$.
We have redone our calculation in six
dimensions and do not find agreement. The slope of $\om_I$ as a function 
of $r_+$ turns out to  be 2.693. Another difference is that the exponents
in black hole critical phenomena are known  to be independent of the
mass of the scalar field. We have checked that the \qn\ frequencies 
of large and intermediate black holes do depend on the mass. One might
expect that if there is a connection between these two phenomena, 
it would apply in the limit of small AdS black holes. However, 
we have seen that the modes of small black holes actually deviate from
the linear relation, so the significance of the asymptotic slope is not clear.
While it is still possible that some deeper connection exists (perhaps
just in four
dimensions and for massless fields) it appears unlikely.

As an aside, we note that if one repeats the calculation of critical
phenomena for spacetimes which are asymptotically AdS, the late time results 
will be quite different. Since energy cannot be lost to infinity,
if one forms a black hole at all, it will eventually grow to absorb all the 
energy of the initial state.

\newsec{Conclusions}

We have computed the scalar quasinormal modes of Schwarzschild-AdS black holes
in four, five, and seven dimensions. These modes govern the late time
decay of a minimally coupled massless scalar field, such as the dilaton.
For large black holes, it is easy to see that these modes must scale with
the black hole temperature $T$.
By the AdS/CFT correspondence, this decay translates
into a timescale for the approach to thermal equilbrium in the CFT, for
large temperatures and perturbations dual to the scalar field. 
The timescale is simply given by the imaginary part of the lowest quasinormal
frequency, $\tau = 1/\om_I$.
From \qntemp, for perturbations with
homogeneous expectation values ($l=0$ modes) these timescales are
$\tau = .0896/T$ for the three dimensional CFT, $\tau =  .116/T$
for the four dimensional super Yang-Mills theory,
and $\tau = .183/T$
for the six dimensional $(0,2)$ theory. 
As we mentioned earlier, these time scales are universal in the sense that
all scalar fields with the same angular dependence will 
decay at this rate.
Perturbations associated with
other linearized supergravity 
fields will decay at different rates, given by their quasinormal mode
frequencies.

Perhaps the most surprising aspect of our analysis are the results for 
intermediate size black holes. For black holes with size of order the AdS
radius, we find that the quasinormal frequencies do not continue to
scale with temperature, but rather scale approximately linearly with horizon
radius. We do not fully understand the implications of this linear relationship
for the dual field theories, but we can make the following comments.
If one considers the field theory at constant temperature, and slowly lowers
the temperature, then one encounters the Hawking-Page transition
\refs{\hapa,\witt}.  At this
point the supergravity description changes from the euclidean black hole to 
a thermal gas in AdS. For these low temperatures, the relaxation time
might still scale with the temperature, but it
cannot be computed by a classical supergravity calculation, and is not
related to quasinormal frequencies.

To interpret the quasinormal frequencies of intermediate size black holes,
we must consider a microcanonical description.
Consider all
states in the CFT with energy equal to the supergravity energy. Most of
these states will be macroscopically indistinguishable, in the sense that
they will all have the same expectation values of the operators dual to 
the supergravity fields. If the only  nonzero expectation value is the
stress energy tensor, the states are described 
on the supergravity side by just 
the black hole. If you perturb one of these CFT states to one which is 
macroscopically slightly different, it will decay to a typical state with
a timescale set by the lowest quasinormal mode. The results in section 4
show that this decay time  
is determined by the size of the black hole in the supergravity description.
Of course the field theory
knows about the black hole size since its entropy is given by the
black hole area. However the fact that, in this range of energy,
the frequency scales linearly with the
radius is puzzling. 

The fact that the quasinormal frequencies do not continue to 
scale with
temperature is also
interesting for the following reason. For a certain range of 
energies,
the  supergravity entropy $S(E)$  is dominated by small (ten
or eleven dimensional) black holes\foot{As we discussed in the previous
section, these black holes can be quantum mechanically stable, since
they are in equilibrium with their Hawking radiation.} \bdhm.
This means that
the effective temperature, 
defined by $dS/dE = 1/T$, has the property that it decreases
as the energy increases, i.e. the specific heat is negative. 
By the AdS/CFT correspondence, the same must be true in the dual CFT.
(This is not a problem since it applies to only a finite range of 
energies.) If the quasinormal modes continued to scale with the temperature,
then this negative specific heat would have dynamical effects. 
Instead, we find that the relaxation time increases monotonically with
decreasing 
energy.
\vskip .5 cm
\centerline{\bf Acknowledgements}

\vskip .2 cm
It is a pleasure to thank P. Brady, M. Choptuik, 
S. Hawking, and B. Schmidt for discussions.
We also wish to thank the Institute for Theoretical Physics, Santa Barbara
where part of 
this work was done.
This work was supported
in part by NSF Grants PHY94-07194 and PHY95-07065.

\vskip 1cm

\centerline{\bf Appendix: Evaluation of \qn\ modes}
\vskip .2 cm

As discussed in section 3, in order to find the \qn\ modes, we need to find
the zeros of $\sum_{n=0}^{\infty} a_n(\om) \, (- x_+)^n$ 
in the complex $\om$ plane.
We compute the \qn\ modes using Mathematica, in the following way.
We search for the zeros, $\om_N$, of
$\psi_{N}(\om) \equiv \sum_{n=0}^{N} a_n(\om) \, (- x_+)^n$ 
by looking for the minima of $|\psi_N|^2$, and checking that the 
value at the minimum is zero, $|\psi_{N}(\om_N)|^2 = 0$.
(In practice, there are numerical errors in the computation,
so the value at the minimum is instead $\sim 10^{-14}$, or smaller.) 
In order to find the correct minimum, we need to specify an
initial guess for $\om_N$.
(This sometimes poses difficulties 
in searching for new modes, and apart from using  analytical or
intuitive understanding as our guide, we are forced to resort to 
trial and error.)  How close to the actual minimum 
one is required to start
depends on the parameters; for the $n=1,\, l=0$ mode of
reasonably-sized black hole, this seldom poses any limitations.

To obtain an accurate estimate of the \qn\ frequencies $\om$,
we typically need to compute on the order of $N=100$ partial sums,
depending on the dimension $d$, the black hole size $r_+$, 
and the mode (i.e.\ $n$ and $l$).
Roughly speaking, at a fixed partial sum $N$, the relative error
in the computed \qn\ frequency grows as $r_+$ decreases, 
and as $l$, $n$, or $d$ increases.

\ifig\fconv{Convergence plot for a five dimensional black hole with $r_+ = 0.6$}
{\epsfxsize=9.5cm \epsfysize=5.5cm \epsfbox{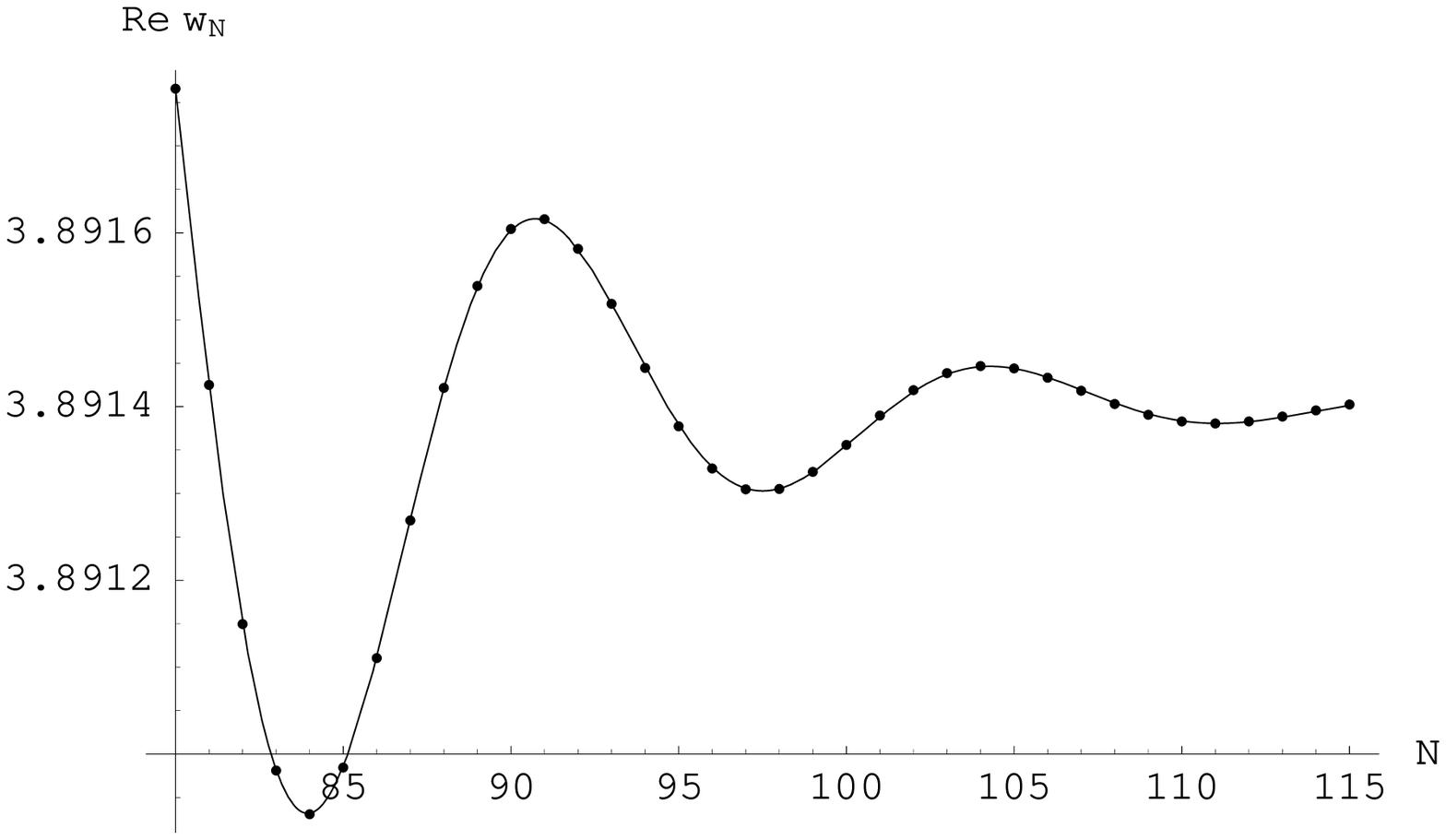}}

The task of determining the mode to the necessary accuracy is 
fortuitously simplified by the fact that the ``convergence curve''
has a surprisingly simple form.  
In particular, once the partial sums have converged to sufficient accuracy,
the variation of $\om_N$
is given by an exponentially damped sine as a function of $N$, 
i.e.\ $ \om_N \sim \om + c\, e^{-N/a}\, \sin(bN +d) $ where $a$ and
$b$ depend on the physical parameters such as $r_+$,
while $c$ and $d$ just depend on which partial sum we start with.
In fact, we can use a fitting algorithm in Mathematica to fit
these convergence curves.
An example is given in \fconv, where the dots represent $Re(\om_N)$ 
for a particular set of parameters ($d=5$, $r_+ = 0.6$, $n=1$, and $l=0$), and 
the solid curve is the corresponding fit.
This simplification allows us to determine the mode with a much higher accuracy
then we would be led to expect from the spread of $\om_N$.
It also allows us to confirm that the numerical errors in the computation
of each $\om_N$ are negligible, since otherwise, one would expect a more
noisy distribution.

Often the quickest way to obtain the \qn\ mode is to simply look for the
minimum of $|\psi_N|^2$ near various initial
guesses for the frequency,
but when that method fails,
 we can also adopt a more systematic approach;  eliminating the 
possibility of occurence of \qn\ modes in a given frequency range.
This may be carried out in a more systematic manner 
due to the fact that $Re(\psi_{N}(\om))$ and $Im(\psi_{N}(\om))$
are conjugate harmonic functions of $\om$, which must satisfy 
the maximum principle.  
Thus, if we find that $\psi_{N}(\om)$ is bounded inside a given
region of the complex $\om$ plane, and either $Re(\psi_{N}(\om))$ or 
$Im(\psi_{N}(\om))$ remains nonzero everywhere on the boundary, 
then $\psi_N$ is necessarily nonzero everywhere inside that region.
This ensures that there can be no \qn\ modes with these frequencies.
We can thus systematically search for the lowest modes by eliminating
the low frequency regions until we find the modes.

Once we find one mode for a given set of parameters, 
continuity of the solution allows us to trace the mode through the
parameter space; that is, we can find $\om$ for nearby values of
$r_+$ and $l$. 
Also, once we know the $n=1$ and $n=2$ modes for a
fixed $r_+$ and $l$, the equal spacing between the modes allows us
to find the higher $n$ modes (provided the numerical errors stay small).

Thus, the procedure for finding $\om(r_+)$ is the following:
We first consider a large black hole, 
where the convergence is good at a low partial sum, e.g.\ $N=40$.
For such a low cut-off $N$ on the partial sum,
we may easily compute $\psi_N(\om, r_+, l)$ in full generality.
We find the desired mode $\om_N(r_+, n, l)$ using the method described
above, and we can check the convergence by comparing 
this result with that obtained for the lower partial sums.
We can now follow the mode to smaller values of $r_+$, until
the convergence becomes too slow, and we need to
compute higher partial sums.  It becomes more practical at this point
to fix all the parameters, and consider $\psi_N$ as a function
of $\om$ only.
This has the numerical advantage of enabling us to compute the partial sums
to much higher $N$; the drawback, of course, is that now we need to recompute
the whole series for each $r_+$.

\vskip 0.5cm

\centerline{\bf Note added:}
{\ninerm \noindent
 After this work was submitted, we have extended our computations of 
the higher $l$ modes for large black holes, up to $l=25$.
From fits of the large $l$ behavior of $\om_I$, we find strong evidence
that the frequencies indeed stay bounded away from zero:
In particular, a fit of the form  
$\om_I(l) = 1.12 + {15.4 \over l \, + \, 11.8} $ has 
$\chi^2 \approx 2 \times 10^{-6}$, 
as opposed to $\chi^2 \approx  10^{-3}$ for a fit with
$\om_I(l \to \infty) \to 0$.}

%
%

\listrefs
\end